\documentclass[a4paper]{amsart}

\RequirePackage{amsmath}
\RequirePackage{bm}
\RequirePackage{amssymb}
\RequirePackage{upref}
\RequirePackage{amsthm}
\RequirePackage{enumerate}
\RequirePackage{pb-diagram}
\RequirePackage{amsfonts}
\RequirePackage[mathscr]{eucal}
\RequirePackage{verbatim}
\RequirePackage{xr}
\RequirePackage{graphicx}
\usepackage{calc}
\usepackage{xspace}
\RequirePackage{color}
\RequirePackage{ifthen}
\RequirePackage{fancybox}
\RequirePackage{mathrsfs}

\newcommand{\cf}{cf.\@\xspace}
\newcommand{\resp}{resp.\@\xspace}

\newcommand{\al}{\alpha}
\newcommand{\bet}{\beta}

\newcommand{\de}{\delta }
\newcommand{\e}{\epsilon}

\newcommand{\f}{\varphi}
\newcommand{\h}{\eta}

\newcommand{\ka}{\kappa}
\newcommand{\lam}{\lambda}
\newcommand{\m}{\mu}
\newcommand{\n}{\nu}
\newcommand{\om}{\omega}

\newcommand{\s}{\sigma}

\newcommand{\z}{\zeta}

\newcommand{\C}{\varGamma}
\newcommand{\D}{\varDelta}

\newcommand{\Lam}{\varLambda}
\newcommand{\Om}{\varOmega}

\newcommand{\so}{{\mc S_0}}

\newcommand{\msp[1]}[1]{\mspace{#1mu}}

\newcommand{\R}[1][n+1]{{\protect\mathbb R}^{#1}}

\newcommand{\Ss}[1][n+1]{{\protect\mathbb S}^{#1}}
\newcommand{\Cc}{{\protect\mathbb C}}

\newcommand{\N}{{\protect\mathbb N}}

\newcommand{\eR}{\stackrel{\lower1ex \hbox{\rule{6.5pt}{0.5pt}}}{\msp[3]\R[]}}
\newcommand{\eN}{\stackrel{\lower1ex \hbox{\rule{6.5pt}{0.5pt}}}{\msp[1]\N}}
\newcommand{\eO}{\stackrel{\lower1ex \hbox{\rule{6pt}{0.5pt}}}{\msc O}}

\DeclareMathOperator{\tr}{tr}

\newcommand\im{\implies}
\newcommand\ra{\rightarrow}

\newcommand\hra{\hookrightarrow}

\newcommand\pde[2]{\frac {\partial#1}{\partial#2}}
 
\newcommand{\un}{\infty}
\newcommand{\A}{\forall}

\newcommand{\uu}{\cup}
\newcommand{\ii}{\cap}
\newcommand{\uuu}{\bigcup}

\newcommand{\uud}{ \stackrel{\lower 1ex \hbox {.}}{\uu}}
\newcommand{\uuud}[1]{ \stackrel{\lower 1ex \hbox {.}}{\uuu_{#1}}}
\newcommand\su{\subset}

\newcommand{\sminus}[1][28]{\raise 0.#1ex\hbox{$\scriptstyle\setminus$}}

\newcommand{\wed}{\wedge}

\newcommand{\abs}[1]{\lvert#1\rvert}

\newcommand{\norm}[1]{\lVert#1\rVert}

\newcommand{\spd}[2]{\protect\langle #1,#2\protect\rangle}

\newcommand{\tup}{\textup}

\newcommand{\mc}{\protect\mathcal}
\newcommand{\msc}{\protect\mathscr}

\newcommand{\tlam}{\tilde\lam}
\newcommand{\tmu}{\tilde\mu}

\providecommand{\bysame}{\makebox[3em]{\hrulefill}\thinspace}

\newcommand{\cq}[1]{\glqq{#1}\grqq\,}

\newcommand{\cqp}[1]{\glqq{\ignorespaces #1\ignorespaces}\grqq}

\newcommand{\bt}{\begin{thm}}
\newcommand{\bl}{\begin{lem}}
\newcommand{\bc}{\begin{cor}}
\newcommand{\bd}{\begin{definition}}
\newcommand{\bpp}{\begin{prop}}
\newcommand{\br}{\begin{rem}}
\newcommand{\bn}{\begin{note}}
\newcommand{\be}{\begin{ex}}
\newcommand{\bes}{\begin{exs}}
\newcommand{\bb}{\begin{example}}
\newcommand{\bbs}{\begin{examples}}
\newcommand{\ba}{\begin{axiom}}
\newcommand{\bas}{\begin{assumption}}

\newcommand{\et}{\end{thm}}
\newcommand{\el}{\end{lem}}
\newcommand{\ec}{\end{cor}}
\newcommand{\ed}{\end{definition}}
\newcommand{\epp}{\end{prop}}
\newcommand{\er}{\end{rem}}
\newcommand{\en}{\end{note}}
\newcommand{\ee}{\end{ex}}
\newcommand{\ees}{\end{exs}}
\newcommand{\eb}{\end{example}}
\newcommand{\ebs}{\end{examples}}
\newcommand{\ea}{\end{axiom}}
\newcommand{\eas}{\end{assumption}}

\newcommand{\bp}{\begin{proof}}
\newcommand{\ep}{\end{proof}}
\newcommand{\eps}{\renewcommand{\qed}{}\end{proof}}

\newcommand{\bal}{\begin{align}}

\newcommand{\bi}[1][1.]{\begin{enumerate}[\upshape #1]}
\newcommand{\bia}[1][(1)]{\begin{enumerate}[\upshape #1]}
\newcommand{\bin}[1][1]{\begin{enumerate}[\upshape\bfseries #1]}
\newcommand{\bir}[1][(i)]{\begin{enumerate}[\upshape #1]}
\newcommand{\bic}[1][(i)]{\begin{enumerate}[\upshape\hspace{2\cma}#1]}
\newcommand{\bis}[2][1.]{\begin{enumerate}[\upshape\hspace{#2\parindent}#1]}
\newcommand{\ei}{\end{enumerate}}

\newcommand\ndots{\raise 0.47ex \hbox {,}\hskip0.06em\cdots %
     \raise 0.47ex \hbox {,}\hskip0.06em} 

\newcommand{\q}{\quad}
\newcommand{\qq}{\qquad}

\newcommand\nd{\noindent}

\newskip\Csmallskipamount                                                
\Csmallskipamount=\smallskipamount
\newskip\Cmedskipamount
\Cmedskipamount=\medskipamount
\newskip\Cbigskipamount
\Cbigskipamount=\bigskipamount

\newcommand\cvs{\vspace\Csmallskipamount}   
\newcommand\cvm{\vspace\Cmedskipamount}

\newskip\csa
\csa=\smallskipamount

\newskip\cma
\cma=\medskipamount

\newskip\cba
\cba=\bigskipamount

\newdimen\spt
\newcommand\citem{\cvs\advance\itemno by
1{(\romannumeral\the\itemno})\hskip3pt}
\newcommand{\bitem}{\cvm\nd\advance\itemno by
1{\bf\the\itemno}\hspace{\cma}}

\newcount\itemno
\newcommand{\las}[1]{\label{S:#1}}

\newcommand{\lae}[1]{\label{E:#1}}
\newcommand{\lat}[1]{\label{T:#1}}
\newcommand{\lal}[1]{\label{L:#1}}

\newcommand{\lac}[1]{\label{C:#1}}

\newcommand{\lar}[1]{\label{R:#1}}

\newcommand{\rs}[1]{Section~\ref{S:#1}}

\newcommand{\rt}[1]{Theorem~\ref{T:#1}}
\newcommand{\rl}[1]{Lemma~\ref{L:#1}}

\newcommand{\rr}[1]{Remark~\ref{R:#1}}

\newcommand{\re}[1]{\eqref{E:#1}}
\newcommand{\frc}[1]{Corollary~\ref{C:#1} on page~\tup{\pageref{C:#1}}}

\newcommand{\frl}[1]{Lemma~\ref{L:#1} on page~\tup{\pageref{L:#1}}}

\newcommand{\fre}[1]{\eqref{E:#1} on page~\tup{\pageref{E:#1}}}

\newskip\thmskip
\thmskip=\parindent

\newskip\hsk
\newenvironment{hinw}{\labelsep=0pt\begin{list}{}{\labelsep=0pt\itemindent=0pt\labelwidth=0pt\leftmargin=\parindent\rightmargin=0pt\partopsep=\cba}%
\item\it\nopagebreak\nopagebreak}%
{\end{list}}

\newcommand\bh{\begin{hinw}}
\newcommand{\eh}{\end{hinw}}

\newtheoremstyle{normal}
  {\cba}
  {\cba}
  {}
  {\thmskip}
  {\bfseries}
  {.}
  {\hsk}
  {}

\newtheoremstyle{abschnitt}
  {\cba}
  {\cba}
  {}
  {\thmskip}
  {\bfseries}
  {.}
  {\hsk}
  {}

\newtheoremstyle{italic}
  {\cba}
  {\cba}
  {\itshape}
  {\thmskip}
  {\bfseries}
  {.}
  {\hsk}
  {}

\newtheoremstyle{aufgaben}
  {\cba}
  {\cba}
  {}
  {}
  {\normalsize\bfseries}
  {.}
  {\hsk}
  {}

\newtheoremstyle{break}
  {\cba}
  {\cba}
  {\itshape}
  {}
  {\bfseries}
  {.}
  {\newline}
  {}

\swapnumbers
\theoremstyle{italic}
\newtheorem{thm}[subsection]{Theorem}
\newtheorem{lem}[subsection]{Lemma}
\newtheorem{prop}[subsection]{Proposition}
\newtheorem{cor}[subsection]{Corollary}

\theoremstyle{normal}
\newtheorem{rem}[subsection]{Remark}
\newtheorem{definition}[subsection]{Definition}
\newtheorem{example}[subsection]{Example}
\newtheorem{examples}[subsection]{Examples}
\newtheorem{ex}[subsection]{Exercise}
\newtheorem{note}[subsection]{}
\newtheorem{axiom}[subsection]{Axiom}
\newtheorem{assumption}[subsection]{Assumption}

\theoremstyle{aufgaben}
\newtheorem{exs}[subsection]{Exercises}

\swapnumbers

\numberwithin{equation}{section}
\numberwithin{figure}{section}

\newenvironment{textequation}[1][0.8]
{\begin{equation}
\begin{aligned}
\begin{minipage}{#1\linewidth}}
{\end{minipage}
\end{aligned}
\end{equation}
\ignorespacesafterend}

\newcommand{\btext}{\begin{textequation}}
\newcommand{\etext}{\end{textequation}}

\def\hinweis{\@startsection{subsection}{2}%
 \z@{0.7\linespacing\@plus 0.5\linespacing}{0.7\linespacing}%
{\normalfont\itshape\indent}}

\newcounter{hours}\newcounter{minutes}
\newcommand{\printtime}{%
\setcounter{hours}{\time/60}%
\setcounter{minutes}{\time-\value{hours}*60}%
\ifthenelse{\value{minutes}<10}{\thehours :0\theminutes}{\thehours:\theminutes}}

\usepackage[german,english]{babel}
\usepackage{graphicx}
\RequirePackage{amsmath}
\RequirePackage{bm}
\RequirePackage{amssymb}
\RequirePackage{upref}
\RequirePackage{amsthm}
\RequirePackage{enumerate}
\RequirePackage{pb-diagram}
\RequirePackage{amsfonts}
\RequirePackage[mathscr]{eucal}
\RequirePackage{verbatim}
\RequirePackage{xr}
\RequirePackage{graphicx}
\usepackage{calc}
\usepackage{xspace}

\makeatletter
\RequirePackage{color}
\newcommand{\ann}[1]{\renewcommand{\@makefnmark}{\mbox{$^{\color{red}{\@thefnmark}}$}}%
\footnote {#1}}
\makeatother

\RequirePackage{upref}
\RequirePackage{amsthm}
\RequirePackage{enumerate}
\usepackage[mathscr]{eucal}

\usepackage{xr-hyper}

\usepackage{calc}

\newlength{\oddsidemarginlength}
\newlength{\topmarginlength}

\newcounter{numberoflines}
\newcounter{tempcc}
\oddsidemargin=\oddsidemarginlength
\evensidemargin=\oddsidemargin
\topmargin=\topmarginlength
\usepackage[colorlinks=true,linkcolor=blue,citecolor=blue,urlcolor=blue]{hyperref}  

\begin{document}

\flushbottom


\title[A partition function for quantized spacetimes]{A partition function for Schwarzschild-AdS and Kerr-AdS black holes and for quantized globally hyperbolic spacetimes with a negative cosmological constant}

\author{Claus Gerhardt}
\address{Ruprecht-Karls-Universit\"at, Institut f\"ur Angewandte Mathematik,
Im Neuenheimer Feld 205, 69120 Heidelberg, Germany}
\email{\href{mailto:gerhardt@math.uni-heidelberg.de}{gerhardt@math.uni-heidelberg.de}}
\urladdr{\href{http://www.math.uni-heidelberg.de/studinfo/gerhardt/}{http://www.math.uni-heidelberg.de/studinfo/gerhardt/}}

%
\date{\today}
%


\begin{abstract} 
We apply quantum statistics to our quantized versions of Schwarzschild-AdS and Kerr-AdS black holes and also to the quantized globally hyperbolic spacetimes having an asymptotically Euclidean Cauchy hypersurface by first proving, for the temporal Hamiltonian $H_0$, that $e^{-\bet H_0}$, $\bet>0$, is of trace class and then, that this result is also valid for the spatial Hamiltonian $H_1$, which has the same eigenvalues but with larger multiplicities. Since the lowest eigenvalue is strictly positive the extension of $e^{-\bet H_1}$ to the corresponding symmetric Fock space is also of trace class and we are thus able to define a partition function $Z$, the operator density $\rho$, the entropy $S$, and the average energy $E$. We prove that $S$ and $E$ tend to infinity if the cosmological constant $\Lam$ tends to $0$ and vanish if $\abs\Lam$ tends to infinity. We also conjecture that $E$ is the source of the dark matter and that the dark energy density is a multiple of the eigenvalue of $\rho$ with respect to the vacuum vector which is $Z^{-1}$.
\end{abstract}

\maketitle

\tableofcontents

\setcounter{section}{0}
\section{Introduction}
In three recent papers we applied our model of quantum gravity to a globally hyperbolic spacetime with an asymptotically Euclidean Cauchy hypersurface \cite{cg:uf4} and to a Schwarzschild-AdS \cite{cg:qbh} \resp Kerr-AdS black hole \cite{cg:qbh2}. In all three cases the quantized model had the same structure, namely, it consisted of special solutions to a wave equation
\begin{equation}\lae{1.1} 
\frac1{32}\frac{n^2}{n-1}\Ddot u
-(n-1) t^{2-\frac4n}\D u-\frac {n}2 t^{2-\frac4n}Ru+nt^2\Lam u=0,
\end{equation}
 in a quantum spacetime
\begin{equation}
N=\R[]_+\times \so,
\end{equation}
where $\so$ is a $n$-dimensional, $n\ge 3$, Cauchy hypersurface of the original spacetime.  The Laplacian and the scalar curvature correspond to the metric $\s_{ij}$ in $\so$, \cf\cite[Theorem 6.9]{cg:qgravity2}, where we derived this wave equation after a canonical quantization process.
The special solutions are a sequence of smooth functions which are a product of temporal and spatial eigenfunctions, where the spatial eigenfunctions are eigendistributions.

In case of the globally hyperbolic spacetime with an asymptotically Euclidean Cauchy hypersurface the solutions to the wave equation can be expressed in the form
\begin{equation}\lae{1.2}
u_{ij}=w_iv_{ij},\qq i\in\N,\; 1\le j\le m\le\un,
\end{equation}
where the $w_i$ are the eigenfunctions of a temporal Hamilton operator $H_0$
\begin{equation}
H_0w_i=\lam_iw_i
\end{equation}
and the $\lam_i$ have multiplicity one such that
\begin{equation}
0<\lam_0<\lam_1<\cdots
\end{equation}
and for each fixed $i$ the at most countably many $v_{ij}$ generate an eigenspace
\begin{equation}
\msc E_{\lam_i}\su \msc S'(\so)
\end{equation}
of a spatial Hamiltonian $H_1$, i.e.,
\begin{equation}
H_1v_{ij}=\lam_iv_{ij}.
\end{equation}
We have
\begin{equation}
v_{ij}\in C^\un(\so)\ii\msc S'(\so).
\end{equation}
In the two remaining cases of the black holes the special solutions are labelled by three indices
\begin{equation}\lae{1.8}
u_{ijk}=w_i\zeta_{ijk}\f_j,
\end{equation}
where the $w_i$ are the same temporal eigenfunctions as before, the $\f_j$ are the eigenfunctions of an elliptic operator $A$ on a smooth compact Riemannian manifold $(M,\s_{ij})$, where  topologically
\begin{equation}
M\simeq \Ss[n-1],
\end{equation}
at least the physically interesting cases, i.e.,
\begin{equation}
A\f_j=\tmu_j \f_j,
\end{equation}
\begin{equation}
\tmu_0<\tmu_1\le\tmu_2\le\cdots
\end{equation}
The $\f_j$ form a mutually orthogonal basis of $L^2(M)$. For a Schwarzschild-AdS black hole we know that
\begin{equation}
\tmu_0\le 0,
\end{equation}
and for a Kerr-AdS black hole this condition can be assured by assuming that the rotational parameter $a$ is small enough such that the scalar curvature of $\s_{ij}$ is positive. Let us emphasize that we considered in \cite{cg:qbh2} Kerr-AdS black holes of odd dimensions
\begin{equation}
\dim N=2m+1,\qq m\ge 2,
\end{equation}
and assumed that all rotational parameters $a_i$ are equal
\begin{equation}
a_i=a\not=0\qq\A\, 1\le i\le m.
\end{equation}
The $\zeta_{ijk}$ are eigendistributions in $\msc S'(\R[])$ satisfying
\begin{equation}
-\zeta''_{ijk}=\om_{ij}^2\zeta_{ijk},\qq k=1,2,
\end{equation}
where
\begin{equation}
\zeta_{ij1}(\tau)=\frac1{\sqrt{2\pi}}e^{i\om_{ij}\tau}
\end{equation}
and
\begin{equation}
\zeta_{ij2}(\tau)=\frac1{\sqrt{2\pi}}e^{-i\om_{ij}\tau},
\end{equation}
where
\begin{equation}
\om_{ij}\ge 0
\end{equation}
is defined by the relation
\begin{equation}\lae{1.19}
\lam_i=\tilde \mu_j+\om_{ij}^2, 
\end{equation}
i.e.,  for any $i\in\N$ we look for all $j$ satisfying
\begin{equation}\lae{1.20}
\tilde\mu_j\le\lam_i
\end{equation}
and then choose $\om_{ij}\ge0$ satisfying \re{1.19}. Let $N_i$ be the set of integers such that the $\tilde \mu_j$ satisfy \re{1.20}, then the smooth functions
\begin{equation}
\zeta_{ijk}\f_j
\end{equation}
are mutually orthogonal in $L^2(M,\s_{ij})$---for fixed $i$ and $k$; note that we only have two different eigendistributions $\z_{ijk}$, if 
\begin{equation}
\om_{ij}>0,
\end{equation}
otherwise we have only one. The eigendistributions $\zeta_{ij1}$ and $\zeta_{ij2}$ are also considered to  be \cq{orthogonal} since their Fourier transforms
\begin{equation}
\hat\zeta_{ijk}=\de_{\pm\om_{ij}}
\end{equation}
have disjoint supports.

Finally, the smooth functions $u_{ijk}$ in \re{1.8} can be considered to be mutually orthogonal since $u_{ijk}$ and $u_{i'j'k'}$ are mutually orthogonal in
\begin{equation}
L^2(\R[]_+,d\mu)\otimes L^2(M),
\end{equation}
where
\begin{equation}
d\mu=t^{2-\frac4n}dt,
\end{equation}
if 
\begin{equation}
\om_{ij}=\om_{i'j'}\q\wed\q k=k'
\end{equation}
and as tempered distributions otherwise.

The $u_{ijk}$ are eigendistributions for both the temporal Hamiltonian $H_0$ as well as for the spatial Hamiltonian $H_1$ with the same eigenvalues $\lam_i$, where now the eigenvalues have finite multiplicities different from $1$ by definition of the eigendistributions and the $u_{ijk}$ also solve the wave equation, since the wave equation can be expressed as
\begin{equation}
\f_0(H_0u-H_1u)=0,
\end{equation}
where $u=u(t,x)$ is a smooth function
\begin{equation}
x\in \so=\R[]\times M
\end{equation}
and
\begin{equation}
\f_0(t)=t^{2-\frac 4n}.
\end{equation}
In \rs{3} we shall prove that we can define an abstract Hilbert space $\mc H$, where the eigendistributions $u_{ijk}$ \resp $u_{ij}$ in \re{1.2} form a basis of mutually orthogonal unit vectors such that the Hamiltonian $H_1$ can be defined on the dense subspace, which is the algebraic span of the basis vectors, as an essentially self-adjoint operator. Let $\tilde H_1$ be its unique self-adjoint extension, namely its closure, then we shall prove that for any $\bet>0$
\begin{equation}\lae{1.30}
e^{-\bet \tilde H_1}
\end{equation}
is of trace class in $\mc H$. In addition $\tilde H_1$ satisfies
\begin{equation}\lae{1.31}
\tilde H_1\ge \lam_0 I,\qq\lam_0>0.
\end{equation}
Let
\begin{equation}
H\equiv d\C(\tilde H_1)
\end{equation}
be the canonical extension of $\tilde H_1$ to the symmetric Fock space 
\begin{equation}
\msc F=\msc F_+(\mc H),
\end{equation}
then
\begin{equation}
e^{-\bet H}
\end{equation}
is of trace class in $\msc F$ because of \re{1.30} and \re{1.31}, \cf \cite[Prop. 5.2.27]{robinson:book2}. Hence we can define the partition function
\begin{equation}
Z=\tr(e^{-\bet H}),
\end{equation}
the density operator
\begin{equation}
\rho=Z^{-1}e^{-\bet H}
\end{equation}
and the von Neumann entropy
\begin{equation}
S=-\tr(\rho\log\rho)=\log Z+\bet E,
\end{equation}
where $E$ is the average energy and $\bet>0$ the inverse temperature
\begin{equation}
\bet=T^{-1}.
\end{equation}
Here is a summary of the results derived in \rs{3}:
\bt \lat{1.1} 
\tup{(i)} Let $\bet_0>0$ be arbitrary, then, for any
\begin{equation}
0<\bet\le\bet_0,
\end{equation}
we have
\begin{equation}
\lim_{\Lam\ra 0}E=\un
\end{equation}
as well as
\begin{equation}
\lim_{\Lam\ra 0}S=\un,
\end{equation}
where the limites are uniform in $\bet$.

\tup{(ii)} Let $\bet_0>0$ be arbitrary, then, for any
\begin{equation}
\bet\ge \bet_0,
\end{equation}
we have
\begin{equation}
\lim_{\abs\Lam\ra\un}E=0
\end{equation}
as well as
\begin{equation}
\lim_{\abs\Lam\ra 0}S=0,
\end{equation}
where the limites are uniform in $\bet$.
\et
The behaviour of $Z$ with respect to $\Lam$ is described in the theorem:
\bt
Let $\bet_0>0$ be arbitrary, then, for any
\begin{equation}
0<\bet\le\bet_0,
\end{equation}
we have
\begin{equation}
\lim_{\Lam\ra 0}Z=\un 
\end{equation}
and for any
\begin{equation}
\bet_0\le\bet
\end{equation}
the relation
\begin{equation}
\lim_{\abs\Lam\ra\un}Z=1
\end{equation}
is valid. The convergence in both limites is uniform in $\bet$.  
\et
\br
The first part of \rt{1.1} reveals that the energy becomes very large for small values of $\abs\Lam$. Since this is the energy obtained by applying quantum statistics to the quantized version
of a black hole or of a globally hyperbolic spacetime---assuming its Cauchy hypersurfaces are asymptotically Euclidean---a small negative cosmological constant might be responsible for the dark matter, where we equate the energy of the quantized universe with matter. As source for the dark energy density we consider the eigenvalue of the density operator $\rho$ with respect to the vacuum vector $\h$
\begin{equation}
\rho\h=Z^{-1}\h,
\end{equation}
i.e., the dark energy density should be proportional to $Z^{-1}$.
\er
\br
Let us also mention that we use Planck units in this paper, i.e.,
\begin{equation}
c=G=\hbar=K_B=1.
\end{equation}
\er
In \rs{4} we also applied quantum statistics to the quantized version of Friedmann universe and proved:
\bt
The results in \rt{3.4}, \rr{3.5} and \rt{3.4} are also valid, if the quantized spacetime $N=N^{n+1}$, $n\ge 3$,  is a Friedmann universe  without matter but with a negative cosmological constant $\Lam$ and with vanishing spatial curvature. The eigenvalues of the spatial Hamiltonian $H_1$ all have multiplicity one.
\et

\section{Trace class estimates}\las{2}
We want to apply quantum statistics to the system described by the  wave equation and its special solutions. Therefore, we need a separable Hilbert space $\mc H$ and a Hamiltonian $H$ such that
\begin{equation}
H\ge\lam_0>0
\end{equation}
and
\begin{equation}
e^{-\bet H},\qq\bet>0,
\end{equation}
is of trace class in $\mc H$.

A natural candidate is the temporal Hamiltonian $H_0$ mentioned in the introduction which corresponds to a generalized eigenvalue problem that has been considered in \cite[Section 4]{cg:uf4}:
Define the bilinear forms
\begin{equation}\lae{2.3}
B(w,\tilde w)=\int_{\R[*]_+}\{\frac 1{32}\frac{n^2}{n-1}\bar w'\tilde w'+n\abs\Lam t^2\bar w\tilde w\}
\end{equation}
and
\begin{equation}\lae{2.4}
K(w,\tilde w)=\int_{\R[*]_+}t^{2-\frac4n}\bar w\tilde w
\end{equation}
in the Sobolev space $\mc H_1$ which is the completion of
\begin{equation}
C^\un_c(\R[*]_+,\Cc)
\end{equation}
in the norm defined by the first bilinear form.

We then look at the generalized eigenvalue problem
\begin{equation}\lae{4.43}
B(w,\f)=\lam K(w,\f)\q\A\,\f\in\mc H_1.
\end{equation}
The following theorem was proved in the former paper.
\bt\lat{4.1}
The eigenvalue problem \re{4.43} has countably many solutions $(w_i,\lam_i)$ such that
\begin{equation}\lae{4.44}
0<\lam_0<\lam_1<\lam_2<\cdots,
\end{equation}
\begin{equation}
\lim\lam_i=\un,
\end{equation}
and
\begin{equation}
K(w_i,w_j)=\de_{ij}.
\end{equation}
The $w_i$ are complete in $\mc H_1$ as well as in $L^2(\R[*]_+)$.
\et

The eigenfunctions $w_i$ solve the ordinary differential equation
\begin{equation}\lae{2.10}
Aw_i=-\frac1{32} \frac{n^2}{n-1}\Ddot w_i+n\abs\Lam t^2w_i=\lam_i t^{2-\frac4n}w_i.
\end{equation}
Let $\f_0=\f_0(t)$ be defined by
\begin{equation}
\f_0(t)=t^{2-\frac4n},
\end{equation}
then the operator
\begin{equation}
\tilde A=\f_0^{-1}A
\end{equation}
is symmetric in 
\begin{equation}
\mc H=L^2(\R[]_+,d\mu),\qq d\mu=\f_0dt,
\end{equation}
and the $w_i$ are eigenfunctions of $\tilde A$
\begin{equation}
\tilde A w_i=\lam_i w_i.
\end{equation}
The equation \re{2.10} is equivalent to
\begin{equation}
\f_0\tilde Aw_i=\lam_i\f_0w_i
\end{equation}
and $\tilde A$ with domain
\begin{equation}
D(\tilde A)=\langle w_i:i\in\N\rangle \su\mc H
\end{equation}
is essentially self-adjoint as will be proved later, \frl{3.1}, in a more general setting.  We denote its unique self-adjoint extension by $H_0$. 

We shall now prove that
\begin{equation}
e^{-\bet H_0},\qq\bet>0,
\end{equation}
is of trace class in $\mc H$.

First, we need two lemmata:
\bl\lal{2.2}
The embedding
\begin{equation}
j:\mc H_1\hra \mc H_0=L^2(\R[]_+,d\tilde\mu),
\end{equation}
where
\begin{equation}
d\tilde\mu =(1+t^2)^{-2}dt,
\end{equation}
is Hilbert-Schmidt.
\el
\bp
Maurin was the first to prove that the embedding
\begin{equation}
H^{m,2}(\Om)\hra L^2(\Om),
\end{equation}
where
\begin{equation}
\Om\su \R[n]
\end{equation}
is a bounded domain, is Hilbert-Schmidt provided
\begin{equation}
m>\frac n2,
\end{equation}
\cf \cite[Theorem 1, p. 336]{maurin:book}. We adapt his proof to the present situation.

Let $w\in \mc H_1$, then, assuming $w$ is real valued,
\begin{equation}\lae{2.23}
\begin{aligned}
\abs{w(t)}^2&=2\int_0^t\dot ww\le 2\int_o^\un\abs{\dot w}^2+\frac12\int_0^\un\abs{w}^2\\
&\le c \norm w_1^2
\end{aligned}
\end{equation}
for all $t>0$, where $\norm\cdot_1$ is the norm in $\mc  H_1$. To derive the last inequality in \re{2.23} we used
\begin{equation}\lae{2.24} 
\int_0^1\abs w^2\le 2\int_0^1\abs{\dot w}^2+\frac12\int_0^1\abs w^2
\end{equation}
which is easily be deduced from the equation in \re{2.23}. The estimate
\begin{equation}
\abs{w(t)}\le c_0\norm w_1\qq\A\, t>0
\end{equation}
is of course also valid for complex valued functions from which infer that, for any $t>0$, the linear form
\begin{equation}
w\ra w(t), \qq w\in\mc H_1,
\end{equation}
is continuous, hence it can be expressed as
\begin{equation}
w(t)=\spd{\f_t}w,
\end{equation}
where
\begin{equation}
\f_t\in\mc H_1
\end{equation}
and
\begin{equation}
\norm{\f_t}_1\le c_0.
\end{equation}
Now, let
\begin{equation}
e_i\in \mc H_1
\end{equation}
be an ONB, then
\begin{equation}
\begin{aligned}
\sum_{i=0}^\un\abs{e_i(t)}^2=\sum_{i=0}^\un\abs{\spd{\f_t}{e_i}}^2=\norm{\f_t}_1^2\le c_0^2.
\end{aligned}
\end{equation}
Integrating this inequality over $\R[]_+$ with respect to $d\tilde\mu$ we infer
\begin{equation}
\sum_{i=0}^\un\int_0^\un\abs{e_i(t)}^2d\tilde\mu\le c^2_0
\end{equation}
completing the proof of the lemma.
\ep
\bl
Let $w_i$ be the eigenfunctions of $H_0$, then there exist positive constants $c$ and $p$ such that
\begin{equation}\lae{2.33}
\norm{w_i}_1\le c\abs{\lam_i}^p\norm{w_i}_0\qq\A\, i\in\N,
\end{equation}
where $\norm\cdot_0$ is the norm in $\mc H_0$.
\el
\bp
We have
\begin{equation}\lae{2.34}
\norm{w_i}_1^2=\lam_i\int_0^\un t^{2-\frac4n}\abs{w_i}^2.
\end{equation}
Let $\e>0$ be arbitrary and define
\begin{equation}
q=\frac2{2-\frac2n}=\frac n{n-1}
\end{equation}
and the conjugate exponent
\begin{equation}
q'=\frac q{q-1}=n,
\end{equation}
then the integral on the right-hand side of \re{2.34} can be estimated from above by
\begin{equation}\lae{2.37}
\frac1q\e^q\int_0^\un\{t^{2-\frac4n} (1+t)^{\frac2n}\}^q\abs{w_i}^2+\frac1{q'}\e^{-q'}.\int_0^\un(1+t)^{-\frac2n q'}\abs{w_i}^2 
\end{equation}
We note that by definition
\begin{equation}
\{t^{2-\frac4n} (1+t)^{\frac2n}\}^q\le(1+t)^2
\end{equation}
and that in view of \re{2.24}
\begin{equation}\lae{2.39}
\int_0^\un (1+t)^2\abs{w_i}^2\le c\norm{w_i}^2_1.
\end{equation}
Combining \re{2.34}, \re{2.37} and \re{2.39} we then infer
\begin{equation}
\norm{w_i}_1^2\le c\frac1q \e^q\lam_i\norm{w_i}_1^2+c\frac1{q'}\e^{-q'}\lam_i\norm{w_i}_0^2
\end{equation}
and deduce further, by choosing $\e$ appropriately, that the result is valid with a different constant $c$.
\ep

We are now ready to prove:
\bt\lat{2.4}
Let $\bet>0$, then the operator
\begin{equation}
e^{-\bet H_0}
\end{equation}
is of trace class in $\mc H$, i.e.,
\begin{equation}
\tr(e^{-\bet H_0})=\sum_{i=0}^\un e^{-\bet\lam_i}=c(\bet)<\un.
\end{equation}
\et
\bp
In view of \rl{2.2} the embedding 
\begin{equation}
j:\mc H_1\hra\mc H_0
\end{equation}
is Hilbert-Schmidt. Let 
\begin{equation}
w_i\in \mc H
\end{equation}
be an ONB of eigenfunctions, then
\begin{equation}
\begin{aligned}
e^{-\bet\lam_i}&=e^{-\bet\lam_i}\norm{w_i}^2= e^{-\bet\lam_i}\lam_i^{-1}\norm{w_i}_1^2\\
&\le e^{-\bet\lam_i}\lam_i^{-1}c\abs{\lam_i}^{2p}\norm{w_i}_0^2,
\end{aligned}
\end{equation}
in view of \re{2.33}, but
\begin{equation}
\begin{aligned}
\norm{w_i}_0^2=\norm{w_i}_1^2\,\norm{\tilde w_i}_0^2=\lam_i\norm{\tilde w_i}_0^2,
\end{aligned}
\end{equation}
where
\begin{equation}
\tilde w_i=w_i \norm{w_i}_1^{-1}
\end{equation}
is an ONB in $\mc H_1$, yielding
\begin{equation}
\sum_{i=0}^\un e^{-\bet\lam_i}\le \tilde c \sum_{i=0}^\un\norm{\tilde w_i}_0^2<\un,
\end{equation}
since $j$ is Hilbert-Schmidt.
\ep
There is also a spatial Hamiltonian $H_1$, which, in the case of the black holes considered, is a direct product of a classical harmonic oscillator in $\R[]$ and an elliptic operator $A$ on a compact, smooth Riemannian manifold $M=M^{n-1}$, $n\ge 3$, with metric $\s_{ij}$, where $A$ has the form
\begin{equation}\lae{2.49}
A\f=-(n-1)\D\f-\frac n2 R\f
\end{equation}
and the Laplacian is the Laplacian in $M$ and $R$ the scalar curvature of the metric. $A$ is self-adjoint with domain
\begin{equation}
D(A)=H^{2,2}(M)\su L^2(M),
\end{equation}
where
\begin{equation}
H^{m,2}(M),\qq m\in M,
\end{equation}
are the usual Sobolev spaces with norm
\begin{equation}
\norm\f_{m,2}^2=\sum_{\abs\al\le m}\int_M\abs{D^\al\f}^2.
\end{equation}
$A$ has a pure point spectrum with countable many eigenvalues $\tilde \mu_j$ with finite multiplicities and mutually orthogonal eigenfunctions $\f_j$ such that
\begin{equation}
\tilde\mu_0<\tilde\mu_1\le\cdots
\end{equation}
and
\begin{equation}
\lim_j\tilde\mu_j=\un.
\end{equation}
We want to prove that
\begin{equation}
e^{-\bet A},\qq\bet >0,
\end{equation}
is of trace class in $L^2(M)$.

The proof of this result will follow  the previous arguments very closely.
\bl\lal{2.5}
Let $m>\frac {n-1}2$, then the embedding
\begin{equation}
j: H^{m,2}(M)\hra L^2(M)
\end{equation}
is Hilbert-Schmidt.
\el
\bp
This result is due to Maurin and its proof is identical with the proof of \rl{2.2} apart from  some obvious modifications.
\ep
We also need the lemma:
\bl
Let $m\in\N$, then there exists $c_m>0$ such that
\begin{equation}\lae{2.57}
\norm\f_{2m,2}^2\le c_m(\norm{A^m\f}^2+\norm\f^2)
\end{equation}
and the bilinear form
\begin{equation}\lae{2.58}
\spd{A^m\f}{A^m\psi}_0+\spd\f\psi_0
\end{equation}
defines an equivalent scalar product in $H^{2m,2}(M)$, where
\begin{equation}
\spd\f\psi_0=\int_M\bar\f\psi.
\end{equation}
\el
\bp
Let
\begin{equation}
f\in H^{m,2}(M)
\end{equation}
and
\begin{equation}
\f\in H^{2,2}(M)
\end{equation}
a solution of
\begin{equation}
A\f=f,
\end{equation}
then it is well-known that
\begin{equation}
\f\in H^{m+2,2}(M)
\end{equation}
and there exists $\tilde c_m$ such that
\begin{equation}
\norm\f_{m+2,2}\le \tilde c_m(\norm f_{m,2}+\norm\f_{0}).
\end{equation}
The constant $\tilde c_m$ also depends on $A$ and $M$. Using this estimate the relation \re{2.57} can be easily proved by induction.
\ep
Now, we are ready to prove:
\bt
Let $A$ be the self-adjoint operator in \re{2.49}, then
\begin{equation}
e^{-\bet A}
\end{equation}
is of trace class in $L^2(M)$ for any $\bet>0$.
\et
\bp
Let $m>\frac {n-1}4$ and equip $H^{2m,2}(M)$ with the scalar product  \re{2.58} such that
\begin{equation}\lae{2.66}
\norm\f_{2m,2}^2=\spd{A^m\f}{A^m\f}_0+\spd\f\f_0,
\end{equation}
then any eigenfunctions $\f_i$, $\f_j$ of $A$ satisfy
\begin{equation}\lae{2.67}
\spd{\f_i}{\f_j}_0=0\im\spd{\f_i}{\f_j}_{2m,2}=0.
\end{equation}
Let $(\f_j)$ be an ONB of eigenfunctions of $A$ in $L^2(M)$ and define
\begin{equation}
\tilde\f_j=\f_i\norm{\f_j}_{2m,2}^{-1},
\end{equation}
then the $\tilde \f_j$ form an ONB in $H^{2m,2}(M)$ and we conclude
\begin{equation}\lae{2.69}
\begin{aligned}
e^{-\bet\tilde\mu_j}&=e^{-\bet\tilde\mu_j}\norm{\f_j}_0^2=e^{-\bet\tilde\mu_j}\norm{\f_j}_{2m,2}^2\,\norm{\tilde\f_j}_0^2\\
&=e^{-\bet\tilde\mu_j}(1+\abs{\tilde\mu_j}^{2m})\norm{\tilde\f_j}_0^2\le c_\bet\norm{\tilde\f_j}_0^2
\end{aligned}
\end{equation}
yielding 
\begin{equation}
\sum_{j=0}^\un e^{-\bet\tilde\mu_j}\le c_\bet\sum_{j=0}^\un \norm{\tilde\f_j}_0^2<\un
\end{equation}
in view of \rl{2.5}.
\ep
With the help of the preceding lemma we can now prove that, in case of the black holes, the spatial Hamiltonian $H_1$ has the property that
\begin{equation}
e^{-\bet H_1}
\end{equation}
is of trace class for all $\bet>0$, where we still have to define an appropriate Hilbert space.

We have
\begin{equation}
H_1v= -\Ddot v -Av,
\end{equation}
where we write $v$ as product
\begin{equation}
v(\tau,x)=\zeta(\tau)\f(x)
\end{equation}
with
\begin{equation}
\tau\in\R[]\q\wed\q x\in M=M^{n-1},
\end{equation}
where $A$ is the differential operator in \re{2.49}. Let $\f_j$ be the eigenfunctions of $A$ with eigenvalues $\tilde\mu_j$, then, for any eigenvalue $\lam_i$ we define
\begin{equation}
N_i=\{j\in\N:\tmu_j\le\lam_i\}
\end{equation}
and $\om_{ij}\ge 0$ such that
\begin{equation}
\om_{ij}^2+\tmu_j=\lam_i.
\end{equation}
Note that
\begin{equation}
0\in N_i\qq\A\, i\in\N,
\end{equation}
since
\begin{equation}
\tmu_0\le 0.
\end{equation}
Let 
\begin{equation}
\zeta_{ijk},\qq k=1,2,
\end{equation}
be the tempered distributions
\begin{equation}
\zeta_{ij1}=\frac1{\sqrt{2\pi}}e^{i\om_{ij}\tau}
\end{equation}
and
\begin{equation}
\zeta_{ij2}=\frac1{\sqrt{2\pi}}e^{-i\om_{ij}\tau},
\end{equation}
where this distinction only occurs for
\begin{equation}
\om_{ij}>0.
\end{equation}
Let $\hat\zeta_{ijk}$ be the Fourier transform of $\zeta_{ijk}$, then
\begin{equation}
\hat\zeta_{ij1}=\de_{\om_{ij}}\q\wed\q \hat\zeta_{ij2}=\de_{-\om_{ij}}
\end{equation}
such that these tempered distributions are considered to be mutually \cqp{orthogonal}. The smooth functions
\begin{equation}
u_{ijk}=\zeta_{ijk}\f_j
\end{equation}
satisfy
\begin{equation}
H_1 u_{ijk}=\lam_i u_{ijk}.
\end{equation}
Label the eigenvalues of $H_1$ including their multiplicities and denote them by $\tlam_i$. Then
\begin{equation}\lae{2.86}
\begin{aligned}
\sum_{i=0}^\un e^{-\bet\tlam_i}\le 2\sum_{i=0}^\un e^{-\bet\lam_i}n(\lam_i)=2\sum_{i=0}^\un e^{-\frac\bet 2\lam_i} e^{-\frac\bet 2\lam_i} n(\lam_i),
\end{aligned}
\end{equation}
where
\begin{equation}
n(\lam_i)=\#N_i.
\end{equation}
\bl\lal{2.8}
Let $\bet_0>0$ be arbitrary, then, for any
\begin{equation}
0<\bet_0\le\bet
\end{equation}
and for any $i\in\N$, the estimate
\begin{equation}\lae{2.89}
e^{-\frac\bet 2\lam_i} n(\lam_i)\le c(\bet)\le c(\bet_0),
\end{equation}
where $c(\bet_0)$ also depends on $A$ but is independent of $i\in \N$.
\el
\bp
Each $N_i$ is the disjoint union
\begin{equation}
N_i'\, \dot\uu \,N_i'',
\end{equation}
where
\begin{equation}
N_i'=\{j\in\N_i:\tmu_j\le0\}
\end{equation}
and $N_i''$ is its complement. The operator $A$ has only finitely many eigenvalues which are non-positive, i.e.,
\begin{equation}
\#N_i'\le n_0\qq\A\,i\in\N,
\end{equation}
hence
\begin{equation}\lae{2.93}
\begin{aligned}
e^{-\frac\bet2\lam_i}n_i(\lam_i)&\le n_0+\sum_{j\in N_i''}e^{-\frac\bet2\lam_i}\le n_0+\sum_{j\in N_i''}e^{-\frac\bet2\tmu_j}\\
&\le n_0+\sum_{j\ge n_0}e^{-\frac\bet2\tmu_j}\\
&=n_0+\sum_{j\ge n_0}e^{-\frac\bet2\tmu_j}(1+\abs\tmu_j^{2m})\, \norm{\tilde\f_j}_0^2\\
&\le n_0+c(\bet)\sum_{j=0}^\un \norm{\tilde\f_j}_0^2<\un,
\end{aligned}
\end{equation}
where we used \re{2.69}. The estimate for the Hilbert-Schmidt norm of the embedding
\begin{equation}
j: H^{m,2}(M)\ra L^2(M)
\end{equation}
depends on $A$, since we used the equivalent norm given in \re{2.66}, and  
\begin{equation}
c(\bet)=\sup_{t>0}e^{-\frac\bet2 t}(1+t^{2m}).
\end{equation}
\ep
\bc\lac{2.9} 
The sum on the left-hand side of \re{2.86} is finite and hence
\begin{equation}
e^{-\bet H_1},\qq\bet>0,
\end{equation}
is of trace class provided we can define a Hilbert space $\mc H$ such that such that the eigendistributions form complete set of eigenvectors in $\mc H$ and $H_1$ is essentially self-adjoint in $\mc H$.
\ec
\bp
The first claim follows immediately by combining \re{2.93} and \rt{2.4}. In \frl{3.1} we shall define the Hilbert space  $\mc H$ and shall prove that $H_1$ is essentially self-adjoint in $\mc H$ and that the eigendistributions form a complete set of eigenvectors in $\mc H$.
\ep
The elliptic operator $A$ also depend on $\Lam$, since the underlying Riemannian metric depends on it. The estimates in the preceding lemma remain valid provided $\abs\Lam$ remains in a compact subset of $\R[]$, since the operator $A$ is then still uniformly elliptic and smooth. However, when
\begin{equation}
\abs\Lam\ra\un,
\end{equation}
then the relation \re{2.57} is no longer valid and a  more sophisticated analysis is necessary to achieve
a corresponding estimate. Let us treat the cases Schwarzschild-AdS and Kerr-AdS black holes separately.

For a Schwarzschild-AdS black hole the operator $A$ can be written in the form
\begin{equation}
A=r_0^{-2}\tilde A,
\end{equation}
where $r_0$ is the black hole radius and
\begin{equation}\lae{2.98} 
\tilde A\f=-(n-1)\tilde\D\f-\frac n2 \tilde R\f.
\end{equation}
Here, the Laplacian and the scalar curvature $\tilde R$ refer to the corresponding quantities of $\Ss[n-1]$ with the standard metric, \cf \cite[equ. (2.12) and (2.14)]{cg:qbh}. The eigenfunctions of $A$ are the eigenfunctions of $\tilde A$. Let $\mu_j$ be the eigenvalues of $\tilde A$ and $\tilde\mu_j$ the eigenvalues of $A$, then
\begin{equation}
\tilde\mu_j=r_0^{-2}\mu_j.
\end{equation}
From the definition of the black hole radius
\begin{equation}
m r_0^{-(n-2)}=1+\frac 2{n(n-1)}\abs\Lam r_0^2
\end{equation}
it is evident that
\begin{equation}
\lim_{\abs\Lam\ra\un}r_0=0
\end{equation}
and also
\begin{equation}\lae{2.102}
\lim_{\abs\Lam\ra\un}\abs\Lam r_0^2=\un,
\end{equation}
though the latter result is only needed when we shall treat the Kerr-AdS case.

We can now prove:
\bl\lal{2.9}
Let $\bet_0$ be arbitrary and $\abs{\Lam_0}$ so large that
\begin{equation}
r_0<1\qq\A\, \abs\Lam>\abs{\Lam_0},
\end{equation}
then for any $i\in\N$, any $\bet\ge \bet_0$ and any $\abs\Lam>\abs{\Lam_0}$
\begin{equation}
e^{-\frac\bet2\lam_i}n(\lam_i)\le c(\bet)\le c(\bet_0), 
\end{equation}
where $c(\bet_0)$ also depends on $\tilde A$ but is independent of $\abs\Lam$ and $i\in\N$.
\el
\bp
We follow the proof of \rl{2.8} but use $\tilde A$ instead of $A$ to define an equivalent norm in $H^{m,2}(M)$, 
\begin{equation}
M=\Ss[n-1].
\end{equation}
Then, we infer, \cf \re{2.93},
\begin{equation}
\begin{aligned}
e^{-\frac\bet2\lam_i}n_i(\lam_i)&\le n_0+\sum_{j\in N_i''}e^{-\frac\bet2\lam_i}\le n_0+\sum_{j\in N_i''}e^{-\frac\bet2\tmu_j}\\
&\le n_0+\sum_{j\ge n_0}e^{-\frac\bet2\tmu_j}\\
&=n_0+\sum_{j\ge n_0}e^{-\frac\bet2\tmu_j}(1+\abs\mu_j^{2m})\, \norm{\tilde\f_j}_0^2\\
&\le n_0+c(\bet)\sum_{j=0}^\un \norm{\tilde\f_j}_0^2<\un.
\end{aligned}
\end{equation}
Here, we used
\begin{equation}
\tilde\mu_j=r_0^{-2}\mu_j>\mu_j>0.
\end{equation}
\ep
Let us now look at Kerr-AdS black holes. In \cite[equ. (2.50)]{cg:qbh2} we described the metric $\s_{ij}$  on $M=\Ss[n-1]$
\begin{equation}
\begin{aligned}
ds_M^2&=\frac{r^2+a^2}{1-a^2 l^2}\big(\de_{ij}d\m^id\m^j+\m_i^2\de_{ij}d\f^i d\f^j\big)\\
&\q+a^2\frac{(1+l^2 r^2) (r^2+a^2)}{r^2 (1-a^2 l^2)^2}\m_i^2\m_j^2d\f^id\f^j.
\end{aligned}
\end{equation}
Here
\begin{equation}
n=2m,\qq m\ge 2,
\end{equation}
and the coordinates $\mu_i$, $1\le i\le m$ are subject to the constraint
\begin{equation}
\sum_{i=1}^m\mu_i^2=1.
\end{equation}
They are the latitudinal coordinates of $\Ss[n-1]$ and the $\f_i$, $ 1\le i\le m$ are the azimuthal coordinates. The metric
\begin{equation}
\de_{ij}d\m^id\m^j+\m_i^2\de_{ij}d\f^i d\f^j
\end{equation}
is the standard metric of $\Ss[n-1]$. The constant $r$ is the radius of the event horizon, $a\not=0$ the rotational parameter and
\begin{equation}
l^2=-\frac1{m(2m-1)}\Lam.
\end{equation}
The relation
\begin{equation}
a^2l^2<1
\end{equation}
is assumed. We also require that $a$ is  small enough such that the scalar curvature $R$ of the metric $\s_{ij}$ is positive. We can write the metric as a conformal metric
\begin{equation}
\s_{ij}=\frac{r^2+a^2}{1-a^2l^2}\tilde\s_{ij}.
\end{equation}
We note that the Schwarzschild-AdS black hole is obtained by setting $a=0$ and that
\begin{equation}
\lim_{a\ra 0}r=r_0,
\end{equation}
the Schwarzschild black hole radius. 

In order to prove the analogue of \rl{2.9} we assume that, when
\begin{equation}
\abs\Lam\ra\un,
\end{equation}
$a$ is supposed so small that
\begin{equation}\lae{2.117} 
\lim_{\abs\Lam\ra\un}\abs\Lam a^2=0
\end{equation}
and
\begin{equation}\lae{2.118} 
\lim_{\abs\Lam\ra \un}\abs\Lam r^2=\un,
\end{equation}
and we emphasize that these assumptions are always satisfied if $a=0$, \cf \re{2.102}. If these are satisfied, then the operator $A$ can be expressed in the form
\begin{equation}
A=\frac{1-a^2l^2}{r^2+a^2}\tilde A,
\end{equation}
where $\tilde A$ converges uniformly in $C^\un(M)$ to the operator $\tilde A$ in \re{2.98}, i.e., for large $\abs\Lam$ $\tilde A$ is uniformly elliptic and smooth such that the number of non-positive eigenvalues $n_0(\tilde A)$ is bounded from above  by the $n_0$ of the limit operator
\begin{equation}
n_0\ge \limsup_{\abs\Lam\ra\un}n_0(\tilde A),
\end{equation}
since $n_0$ is upper semi-continuous as it is well-known.
\bl\lal{2.10}
Under the assumptions \re{2.117} and \re{2.118} the results of \rl{2.9} are also valid for the Kerr-AdS black hole, i.e., there exists $\abs{\Lam_0}>0$ such that for all
\begin{equation}
\abs\Lam>\abs{\Lam_0}
\end{equation}
and for any $\bet$ satisfying
\begin{equation}
0<\bet_0\le\bet,
\end{equation}
where $\bet_0$ is arbitrary,
\begin{equation}
e^{-\frac\bet2\lam_i}n(\lam_i)\le c(\bet_0)
\end{equation}
uniformly in $i\in\N$, $\abs\Lam$ and $\bet$. 
\el
\bp
The proof is identical to the proof of \rl{2.9} by using the fact that the special $H^{m,2}(M)$ norm
\begin{equation}
\spd{\tilde A^m\f}{\tilde A^m\f}_0+\spd\f\f_0,
\end{equation}
with different $m$ than used to express the dimension of $M$, is uniformly equivalent to the standard $H^{m,2}(M)$ norm, hence the Hilbert-Schmidt norm of the embedding
\begin{equation}
j:H^{m,2}(M)\hra L^2(M)
\end{equation}
is uniformly bounded. We also relied on
\begin{equation}
\tilde\mu_j=\frac{1-a^2l^2}{r^2+a^2}\mu_j>\mu_j>0
\end{equation}
for $j\in N_i''$.
\ep
Finally, let us derive the last result in this section.
\bl\lal{2.12}
Let $\lam_i$ be the temporal eigenvalues depending on $\Lam$ and let $\bar\lam_i$ be the corresponding eigenvalues for
\begin{equation}
\abs\Lam=1,
\end{equation}
then
\begin{equation}\lae{2.72}
\lam_i=\bar\lam_i\abs\Lam^\frac{n-1}n.
\end{equation}
\el
\bp
Let $B$ and $K$ be the bilinear forms defined in \re{2.3} \resp \re{2.4}, where $B$ corresponds to the cosmological constant $\Lam$ and let $B_1$ be the form with respect to the value
\begin{equation}
\abs\Lam=1.
\end{equation}
Moreover, let us denote the corresponding quadratic forms by the same symbols, then we have
\begin{equation}\lae{2.74}
\frac{B(\f)}{K(\f)}=\abs\Lam^\frac{n-1}n\frac{B_1(\f)}{K(\f)}\qq\A\,0\not=\f\in C^\un_c(\R[]_+).
\end{equation}
To prove \re{2.74} we introduce a new integration variable $\tau$ on the left-hand side
\begin{equation}
t=\mu\tau,\qq\mu>0,
\end{equation}
to conclude
\begin{equation}
\frac{B(\f)}{K(\f)}=\mu^{-4\frac{n-1}n}\frac{B_1(\f)}{K(\f)}\qq\A\,0\not=\f\in C^\un_c(\R[]_+).
\end{equation}
provided
\begin{equation}
\mu=\abs\Lam^{-\frac14}.
\end{equation}
The relation \re{2.74} immediately implies \re{2.72}.
\ep

\section{The partition function}\las{3}
We first define the partition function for the black holes and shall later  show that the definitions and results are also applicable in case of the quantized globally hyperbolic spacetimes with a negative cosmological constant and asymptotically Euclidean Cauchy hypersurfaces.

We define the partition function by using the spatial Hamiltonian $H_1$ of the quantized black holes, Kerr or Schwarzschild, which is now defined in the separable Hilbert space $\mc H$ generated by the eigendistributions
\begin{equation}
u_{ijk}=w_i\zeta_{ijk}\f_j
\end{equation}
which are smooth functions satisfying the eigenvalue equations
\begin{equation}
H_1u_{ijk}=\lam_i u_{ijk}
\end{equation}
as well as
\begin{equation}
H_0 u_{ijk}=\lam_i u_{ijk},
\end{equation}
where $H_0$ is the temporal Hamiltonian. 

In order to explain how the eigendistributions can generate a Hilbert space let us relabel the eigenfunctions and the eigenvalues by $(u_i,\tlam_i)$  such that
\begin{equation}
H_1 u_i=\tlam_i u_i
\end{equation}
and
\begin{equation}
H_0 u_i=\tlam_i u_i,
\end{equation}
i.e., the multiplicities of the eigenvalues are now included in the labelling and the ordering is no longer strict
\begin{equation}
\tlam_0\le \tlam_1\le \tlam_2\le \cdots.
\end{equation}
To define the Hilbert space $\mc H$ we simply declare that the eigendistributions are mutually orthogonal unit eigenvectors, hence defining a scalar product in the  complex vector space $\mc H'$ spanned by these eigenvectors. We define the Hilbert space $\mc H$ to be its completion.

\bl\lal{3.1}
The linear operator $H_1$ with domain $\mc H'$ is essentially self-adjoint  in $\mc H$. Let $\bar H_1$ be its closure, then the only eigenvectors of $\bar H_1$ are those of $H_1$.
\el
\bp
$H_1$ is obviously densely defined, symmetric and bounded from below
\begin{equation}
H_1\ge \tlam_0 I>0.
\end{equation}
Since $\tlam_0>0$, the eigenvectors also span $R(H_1)$, i.e., $R(H_1)$ is dense. Let 
\begin{equation}
w\in \mc H
\end{equation}
be arbitrary, and let 
\begin{equation}
H_1v_i\in R(H_1)
\end{equation}
be a sequence converging to $w$, then $v_i$ is a Cauchy sequence, because
\begin{equation}
\tilde\lam_0\norm{v_i-v_j}^2\le \spd{H_1v_i-H_1v_j}{v_i-v_j}\le \norm{H_1v_i-H_1v_j}\,\norm{v_i-v_j},
\end{equation}
hence
\begin{equation}
R(\bar H_1)=\mc H
\end{equation}
and $\bar H_1$ is the unique s.a. extension of $H_1$.

It remains to prove that $\bar H_1$ has no additional eigenvectors. Thus, let $u$ be an eigenvector of $\bar H_1$ with eigenvalue $\lam$
\begin{equation}
\bar H_1 u=\lam u,
\end{equation}
and let
\begin{equation}
E(\tlam_i)\su \mc H',\qq i\in \N,
\end{equation}
be the eigenspaces of $H_1$. Let us first assume that there exists $j$ such that
\begin{equation}
\lam=\tlam_j,
\end{equation}
but
\begin{equation}
u\notin E(\tlam_j).
\end{equation}
Without loss of generality we may assume
\begin{equation}
u\in E(\tlam_j)^\perp.
\end{equation}
However, this leads to a contradiction, since then
\begin{equation}\lae{3.15}
u\in E(\tlam_i)^\perp\qq\A\, i\in\N,
\end{equation}
and hence
\begin{equation}
u\in \mc H'^\perp
\end{equation}
which implies $u=0$.

Thus, let us assume
\begin{equation}
\lam\not=\tlam_i\qq\A\, i\in \N,
\end{equation}
but then \re{3.15} is again valid leading to the known contradiction.
\ep
\br
In the following we shall write $H_1$ instead of $\bar H_1$.
\er
\bl
For any $\bet>0$ the operator
\begin{equation}
e^{-\bet H_1}
\end{equation}
is of trace class in $\mc H$. Let
\begin{equation}
\msc F\equiv\msc F_+(\mc H)
\end{equation}
be the symmetric Fock space generated by $\mc H$ and let
\begin{equation}
H=d\C(H_1)
\end{equation}
be the canonical extension of $H_1$ to $\msc F$. Then
\begin{equation}
e^{-\bet H}
\end{equation}
is also of trace class in $\msc F$
\begin{equation}\lae{3.24}
\tr(e^{-\bet H})=\prod_{i=0}^\un(1-e^{-\bet \tlam_i})^{-1}<\un.
\end{equation}
\el
\bp
The first part of the lemma has already been proved in \frc{2.9}. This property can now be rephrased as
\begin{equation}\lae{3.25}
\tr(e^{-\bet H_1})=\sum_{i=0}^\un e^{-\bet \tlam_i}<\un.
\end{equation}
The second assertion is well known, since
\begin{equation}\lae{3.26}
H_1\ge \tlam_0 I>0,
\end{equation}
and the properties \re{3.25} and \re{3.26} imply \re{3.24}, \cf \cite[Proposition 5.2.7]{robinson:book2} and \cite[Volume II, p. 868]{honegger:fock}, where the equation \re{3.24} is also proved.
\ep
We then define the partition function $Z$ by
\begin{equation}
Z=\tr(e^{-\bet H})=\prod_{i=0}^\un (1-e^{-\bet\tlam_i})^{-1}
\end{equation}
and the density operator $\rho$ in $\msc F$ by
\begin{equation}
\rho=Z^{-1}e^{-\bet H}
\end{equation}
such that
\begin{equation}
\tr \rho=1.
\end{equation}

The von Neumann entropy $S$ is then defined by
\begin{equation}
\begin{aligned}
S&=-\tr(\rho\log \rho)\\
&=\log Z+\bet Z^{-1}\tr (He^{-\bet H})\\
&=\log Z-\bet\pde{\log Z}\bet\\
&\equiv \log Z +\bet E,
\end{aligned}
\end{equation}
where $E$ is the average energy
\begin{equation}
E=\tr (H\rho).
\end{equation}
$E$ can be expressed in the form
\begin{equation}
E=\sum_{i=0}^\un \frac{\tlam_i}{e^{\bet\tlam_i}-1}.
\end{equation}
Here, we also set the Boltzmann constant
\begin{equation}
K_B=1.
\end{equation}
The parameter $\bet$ is supposed to be the inverse of the absolute temperature $T$
\begin{equation}
\bet=T^{-1}.
\end{equation}
In view of \frl{2.12} we can write the eigenvalues $\lam_i$ in the form
\begin{equation}\lae{3.35}
\lam_i=\bar\lam_i \abs\Lam^\frac{n-1}n,
\end{equation}
where $\bar\lam_i$ are the eigenvalues corresponding to $\abs\Lam=1$. Hence, $Z$, $S$, and $E$ can also be looked at as functions depending on $\bet$ and $\Lam$, or more conveniently, on $(\bet,\tau)$, where
\begin{equation}
\tau=\abs\Lam^\frac{n-1}n,
\end{equation} 
since the $\tlam_i$ can also be expressed as
\begin{equation}\lae{3.37}
\tlam_i=\lam_j=\bar\lam_j\abs\Lam^\frac{n-1}n,
\end{equation}
where $j$ is different from $i$
\begin{equation}
j\le i,
\end{equation}
because of the multiplicities of $\tlam_i$. Let emphasize that the multiplicities also depend on $\Lam$, hence it is best to simply note that
\begin{equation}
\tlam_0=\lam_0=\bar\lam_0\abs\Lam^\frac{n-1}n
\end{equation}
and that the $\tlam_i$ are ordered. We shall never use the relation \re{3.37} explicitly in the proofs of the subsequent theorems and lemmata referring to \re{3.35} instead.

\bt\lat{3.4}
\tup{(i)} Let $\bet_0>0$ be arbitrary, then, for any
\begin{equation}
0<\bet\le\bet_0,
\end{equation}
we have
\begin{equation}
\lim_{\Lam\ra 0}E=\un
\end{equation}
as well as
\begin{equation}\lae{3.42}
\lim_{\Lam\ra 0}S=\un,
\end{equation}
where the limites are uniform in $\bet$.

\tup{(ii)} Let $\bet_0>0$ be arbitrary, then, for any
\begin{equation}\lae{3.43}
\bet\ge \bet_0,
\end{equation}
we have
\begin{equation}
\lim_{\abs\Lam\ra\un}E=0
\end{equation}
as well as
\begin{equation}
\lim_{\abs\Lam\ra 0}S=0,
\end{equation}
where the limites are uniform in $\bet$.
\et
\bp
\cq{(i)} We first observe that
\begin{equation}
E=\sum_{i=0}^\un\frac{\tlam_i}{e^{\bet\tlam_i}-1}\ge \sum_{i=0}^\un\frac{\lam_i}{e^{\bet\lam_i}-1}
\end{equation}
Now, let $m\in\N$ be arbitrary, then
\begin{equation}
E\ge \sum_{i=0}^m\frac{\lam_i}{e^{\bet\lam_i}-1}=\sum_{i=0}^m\frac{\bar\lam_i\tau}{e^{\bet\lam_i\tau}-1}
\end{equation}
and
\begin{equation}
\begin{aligned}
\liminf_{\tau\ra 0}E&\ge \lim_{\tau\ra0}\sum_{i=0}^m\frac{\bar\lam_i\tau}{e^{\bet\lam_i\tau}-1}\\
&=(m+1)\bet^{-1}\ge (m+1)\bet_0^{-1}
\end{aligned}
\end{equation}
yielding
\begin{equation}
\lim_{\Lam\ra0}E=\un
\end{equation}
uniformly in $\bet$.

Since $Z\ge 1$, the relation \re{3.42} follows as well.

\cvm
\cq{(ii)} We estimate $E$ from above by
\begin{equation}
\begin{aligned}
E&=\sum_{i=0}^\un\frac{\tlam_i e^{-\bet\tlam_i}}{1-e^{-\bet\tlam_i}}=\sum_{i=0}^\un\tlam_ie^{-\frac\bet2\tilde\lam_i}e^{-\frac\bet2\tlam_i}(1-e^{-\bet\tlam_i})^{-1}\\
&\le (1-e^{-\bet_0\tlam_0})^{-1}c(\bet_0)\sum_{i=0}^\un e^{-\frac\bet2\tlam_i},
\end{aligned}
\end{equation}
where we used \re{3.43} and
\begin{equation}
\tilde\lam_i e^{-\frac\bet2\tlam_i}\le \sup_{t>0}te^{-\frac\bet2t}=c(\bet)\le c(\bet_0).
\end{equation}
Furthermore, we know that
\begin{equation}
\begin{aligned}
\sum_{i=0}^\un e^{-\frac\bet2\tlam_i}&\le \tilde c(\bet)\sum_{i=0}^\un e^{-\frac\bet4\lam_i}\\
&\le \tilde c(\bet_0)\sum_{i=0}^\un e^{-\frac{\bet_0}4\lam_i},
\end{aligned}
\end{equation}
\cf \frl{2.9} and \frl{2.10}, hence we obtain
\begin{equation}
E\le (1-e^{-\bet_0\bar\lam_0\tau})^{-1}c(\bet_0)\tilde c(\bet_0)\sum_{i=0}^\un e^{-\frac\bet4\bar\lam_i\tau}
\end{equation}
deducing further
\begin{equation}\lae{3.54}
\limsup_{\tau\ra \un}E\le c(\bet_0)\tilde c(\bet_0)\lim_{\tau\ra\un}\sum_{i=0}^\un e^{-\frac\bet4\bar\lam_i\tau}=0
\end{equation}
uniformly in  $\bet$ and hence
\begin{equation}
\lim_{\tau\ra \un}E=0.
\end{equation}

It remains to prove that $S$ vanishes in the limit. We have
\begin{equation}
\begin{aligned}
Z&=\prod_{i=0}^\un (1-e^{-\bet\tlam_i})^{-1}=\prod_{i=0}^\un(1+e^{-\bet\tlam_i}(1-e^{-\bet\tlam_i})^{-1})\\
&\le\exp\{(1-e^{\bet_0\tlam_0})^{-1}\sum_{i=0}^\un e^{-\bet\tlam_i}\},
\end{aligned}
\end{equation}
where we used the inequality
\begin{equation}
\log(1+t)\le t\qq\A\, t\ge 0
\end{equation}
in the last step.

Applying then the arguments preceding the inequality \re{3.54} we conclude
\begin{equation}\lae{3.58}
\lim_{\tau\ra\un}Z=1
\end{equation}
uniformly in $\bet$. 
\ep
\br\lar{3.5}
The first part of the preceding theorem reveals that the energy becomes very large for small values of $\abs\Lam$. Since this is the energy obtained by applying quantum statistics to the quantized version
of a black hole or of a globally hyperbolic spacetime---assuming its Cauchy hypersurfaces are asymptotically Euclidean---a small negative cosmological constant might be responsible for the dark matter, where we equate the energy of the quantized universe with matter. As source for the dark energy density we conjecture that the dark energy density should be proportional to the eigenvalue of the density operator $\rho$ with respect to the vacuum vector $\h$
\begin{equation}
\rho\h=Z^{-1}\h,
\end{equation}
which is  $Z^{-1}$.
\er
The behaviour of $Z$ with respect to $\Lam$ is described in the theorem:
\bt\lat{3.6}
Let $\bet_0>0$ be arbitrary, then, for any
\begin{equation}\lae{3.60}
0<\bet\le\bet_0,
\end{equation}
we have
\begin{equation}
\lim_{\Lam\ra 0}Z=\un
\end{equation}
and for any
\begin{equation}
\bet_0\le\bet
\end{equation}
the relation
\begin{equation}\lae{3.63}
\lim_{\abs\Lam\ra\un}Z=1
\end{equation}
is valid. The convergence in both limites is uniform in $\bet$. 
\et
\bp
\cq{\re{3.60}} Let $m\in\N$ be arbitrary, then
\begin{equation}
\begin{aligned}
Z&\ge \prod_{i=0}^\un (1-e^{-\bet\lam_i})^{-1}=\prod_{i=0}^\un (1-e^{-\bet\bar\lam_i \tau})^{-1}\\
&\ge \prod_{i=0}^m(1-e^{-\bet_0\bar\lam_i \tau})^{-1}
\end{aligned}
\end{equation}
and we infer
\begin{equation}
\lim_{\tau\ra0}Z=\liminf_{\tau\ra0}Z=\un.
\end{equation}

\cvm
\cq{\re{3.63}} This limit relation has already been proved in \re{3.58}. 
\ep
Let us now consider the quantized globally hyperbolic spacetimes with an asymptotically Euclidean Cauchy hypersurface. The eigenspaces
\begin{equation}
\msc E_{\lam_i}\su \msc S'(\so)
\end{equation}
of $H_1$ are separable but they are in general not finite dimensional as can be seen by the following counterexample
\begin{equation}
H_1=-\D
\end{equation}
in $\R[n]$. The eigenspaces
\begin{equation}
\msc E_{\lam_i},\qq \lam_i>0,
\end{equation}
contain the tempered distributions
\begin{equation}\lae{3.69}
e^{i\spd kx},\qq k\in \Ss[n-1]_{\lam_i}.
\end{equation}
As a Hamel basis they generate a vector space the dimension of which is equal to the cardinality of $\Ss[n-1]$. Of course, as a Schauder basis the functions with
\begin{equation}
k\in D\su\Ss[n-1]_{\lam_i},
\end{equation}
where $D$ is countable and dense, generate a dense subspace.

This example indicates that not all eigendistributions of $H_1$ might be physically relevant. Contrary to the cases of the black holes, where the selection of eigenvectors and eigendistributions was a natural process, only the temporal eigenvectors are naturally selected in the present situation and of course at least one matching spatial eigendistribution to obtain a solution of the wave equation. Hence, we could  use $H_0$ to define the partition function. However, we believe this choice would be too restrictive, and we shall instead stipulate that we only pick at most 
\begin{equation}
c\abs{\lam_i}^p
\end{equation}
spatial eigendistributions in $\msc E_{\lam_i}$, where $c$ and $p$ are arbitrary but fixed constants, i.e., we assume that
\begin{equation}\lae{3.72}
n(\lam_i)\le c\abs{\lam_i}^p\qq\A\, i\in\N.
\end{equation}
With this assumption it becomes evident that the results and conjectures of \rt{3.4}, \rr{3.5} and \rt{3.6} are also valid in case of  globally hyperbolic spacetimes with asymptotically Euclidean hypersurfaces.

\section{The Friedmann universes with negative cosmological constants}\las{4}
In \cite[Remark 6.11]{cg:qgravity2} we observed that, if the Cauchy hypersurface $\so$  is a space of constant curvature and if the wave equation  \fre{1.1} is only considered  for functions $u$ which do not depend on $x$,  then this equation is identical to the equation obtained by quantizing the Hamilton constraint in a Friedman universe without matter but including a cosmological constant. The equation is then the ODE
\begin{equation}
\frac1{32} \frac {n^2}{n-1} \Ddot u-\frac n2 R t^{2-\frac 4n}u+nt^2\Lam u=0,\qq 0<t<\un,
\end{equation}
where $R$ is the scalar curvature of $\so$. We cannot apply our previous arguments to the solutions of this ODE. However, if we consider instead the more general equation \re{1.1}, where $u$ is also allowed to depend on $x$, which certainly is more general and accurate, then the previous arguments can be applied if the curvature $\tilde\ka$ of $\so$ vanishes
\begin{equation}
\tilde\ka=0.
\end{equation}The scalar curvature, which is equal to
\begin{equation}
R=n(n-1)\tilde\ka,
\end{equation}
 then vanishes too and 
\begin{equation}
\so=\R[n].
\end{equation}
We are now in the situation which we analyzed at the end of the previous section, where now the spatial Hamiltonian is
\begin{equation}
H_1=-\frac n2\D
\end{equation}
and some spatial eigendistributions are  shown in \fre{3.69}. However, since we consider the quantized version of a Friedmann universe we shall look for radially symmetric eigendistributions, i.e., we look for smooth functions $v=v(x)$ satisfying
\begin{equation}
v(x)=\f(r)
\end{equation}
such that
\begin{equation}\lae{4.7} 
\D v=\Ddot\f+(n-1)r^{-1}\dot\f=-\mu^2\f\q\text{in}\q r>0,
\end{equation}
where $\mu>0$. Obviously, it is sufficient to assume $\mu=1$, because, if $\f$ is an eigenfunction for $\mu=1$, then
\begin{equation}
\tilde\f(r)=\f(\mu r)
\end{equation}
is an eigenfunction for the eigenvalue $\mu^2$. Therefore, let us choose $\mu=1$. 

We shall express the solution $\f$ with help of a Bessel function $J_\n$. Let $\psi$ be a solution of the Bessel equation
\begin{equation}\lae{4.9}
\Ddot\psi+r^{-1}\dot\psi+(1-r^{-2}\nu^2)\psi=0,
\end{equation}
where
\begin{equation}
\nu=\frac{n-2}2,
\end{equation}
then the function
\begin{equation}
\f(r)=r^{-\nu}\psi
\end{equation}
satisfies
\begin{equation}
r\Ddot\f+(2\nu+1)\dot\f+r\f=0,
\end{equation}
which is equivalent to \re{4.7} with $\mu=1$. The Bessel equation \re{4.9} has the two independent solutions $J_\nu$ and $Y_\nu$, the Bessel functions of first kind \resp of second kind. It is well known that the functions
\begin{equation}
r^{-\nu}J_\nu
\end{equation}
can be expressed as a power series in the variable $r^2$, \cf \cite[equ.~(21), p.~420]{courant-hilbert-I}, i.e., the function
\begin{equation}
v(x)=\f(r)=r^{-\nu}J_\nu
\end{equation}
is smooth in $\R[n]$, while the functions
\begin{equation}
r^{-\nu}Y_\nu
\end{equation}
have a singularity in $r=0$. Hence, there exists exactly one smooth radially symmetric solution $v$ of the eigenvalue equation
\begin{equation}
-\D v=\lam^2 v,\qq \lam>0,
\end{equation}
which is given by
\begin{equation}
v=(\lam r)^{-\nu}J_\nu(\lam r).
\end{equation}
This solution also vanishes at infinity, hence it is uniformly bounded and a tempered distribution. 

A solution of the wave equation \fre{1.1}, in case of a quantized Friedmann universe, is therefore given by a sequence
\begin{equation}
u_i=w_i(t)v_i(x),\qq i\in\N,
\end{equation}
where $w_i$ is a temporal eigenfunction and $v_i$ a spatial eigenfunction. The $u_i$ are also eigenfunctions for the temporal Hamiltonian as well as for the spatial Hamiltonian. Each eigenvalue has multiplicity one. We have therefore proved:

\bt
The results in \rt{3.4}, \rr{3.5} and \rt{3.4} are also valid, if the quantized spacetime $N=N^{n+1}$, $n\ge 3$,  is a Friedmann universe  without matter but with a negative cosmological constant $\Lam$ and with vanishing spatial curvature. The eigenvalues of the spatial Hamiltonian $H_1$ all have multiplicity one.
\et

\bibliographystyle{hamsplain}
\providecommand{\bysame}{\leavevmode\hbox to3em{\hrulefill}\thinspace}
\providecommand{\href}[2]{#2}



\end{document}